# Investigating the Product Profiles and Structural Relationships of New Levansucrases with Conventional and Non-Conventional Substrates

Andrea Hill [1], Salwa Karboune [1,*], Tarun J. Narwani [2] and Alexandre G. de Brevern [2]

1 Department of Food Science, McGill University, 21 111 Lakeshore, Ste-Anne-de-Bellevue, QC H9X 3V9, Canada; andrea.hill3@mail.mcgill.ca
2 INSERM, UMR-S 1134, Dynamique des Structures et Interactions des Macromolécules Biologiques (DSIMB), Univ Paris, Institut National de la Transfusion Sanguine (INTS), 75015 Paris, France; tjrnarwani@gmail.com (T.J.N.); Alexandre.debrevern@univ-paris-diderot.fr (A.G.d.B.)
* Correspondence: salwa.karboune@mcgill.ca

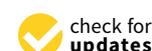

**Abstract:** The synthesis of complex oligosaccharides is desired for their potential as prebiotics, and their role in the pharmaceutical and food industry. Levansucrase (LS, EC 2.4.1.10), a fructosyl-transferase, can catalyze the synthesis of these compounds. LS acquires a fructosyl residue from a donor molecule and performs a non-Lenoir transfer to an acceptor molecule, via β-(2→6)-glycosidic linkages. Genome mining was used to uncover new LS enzymes with increased transfructosylating activity and wider acceptor promiscuity, with an initial screening revealing five LS enzymes. The product profiles and activities of these enzymes were examined after their incubation with sucrose. Alternate acceptor molecules were also incubated with the enzymes to study their consumption. LSs from *Gluconobacter oxydans* and *Novosphingobium aromaticivorans* synthesized fructooligosaccharides (FOSs) with up to 13 units in length. Alignment of their amino acid sequences and substrate docking with homology models identified structural elements causing differences in their product spectra. Raffinose, over sucrose, was the preferred donor molecule for the LS from *Vibrio natriegens, N. aromaticivorans*, and *Paraburkholderia graminis*. The LSs examined were found to have wide acceptor promiscuity, utilizing monosaccharides, disaccharides, and two alcohols to a high degree.

**Keywords:** levansucrase; fructooligosaccharide; fructosylation; prebiotic; alcohol; acceptor specificity; homology modeling

## 1. Introduction

Biosynthetic routes for the synthesis of novel carbohydrates is an attractive course. Enzymatic glycosylation reactions can proceed via a regio- and stereo-selective manner, unlike chemical synthesis, which requires the use of protecting groups [1]. Levansucrase (EC 2.4.1.10, LS) is a β-fructosyl transferase capable of catalyzing the non-Lelior-type transfructosylation reaction generating prebiotic fructooligosaccharides (FOSs) and the fructan polysaccharide levan. Levan is composed of β-(2-6)-linked fructosyl residues. It has its own functional properties for food and pharmaceutical industries and can also be hydrolyzed to FOSs [2–5]. FOSs have interesting properties as prebiotics, having bifidogenic and anticarcinogenic effects, which make their synthesis desirable [6–8].

LS belongs to glycosyl hydrolase family 68 and is expressed by both Gram-positive and Gram-negative bacteria. Belonging to clan GH-J, along with family GH-32, their structure contains 5-fold β-propeller topology with four anti-parallel strands [9]. LS catalyzes these reactions via a "ping-pong" or double displacement mechanism. This action occurs within a deep negative cavity on the enzyme. The fructofuranosyl group is donated by an acceptor molecule, such as sucrose, raffinose, or stachyose [10]. An enzyme intermediate is formed with the fructosyl group, with the remainder of the donor molecule released. LS can catalyze different reactions, depending upon the acceptor of the fructosyl group. Hydrolysis of the donor molecule occurs when the acceptor is water, releasing fructose. Transfructosylation occurs when the fructosyl group is transferred to the acceptor molecule, typically via a β-(2-6)-glycosidic linkage. Transfructosylation can be classified as either oligomerization or polymerization [11]. LSs can be characterized as predominately having a processive reaction or a disproportionate reaction depending upon the microbial source [12]. In a processive reaction, the enzyme has subsites, which hold onto the growing transfructosylated product, and produce levan and larger FOSs [13]. In a disproportionate reaction, the LS lacks affinity for the product, releasing it after performing the transfructosylation reaction, resulting in shorter FOS products [11,14].

LS can utilize other non-sucrose-derived acceptor molecules due to the flexibility of the +1 subsite [15]. LS has been shown to use alternate monosaccharides to create sucrose analogues [16–18]. The sucrose analogues can in turn act as acceptor molecules for further transfructosylation reactions, producing hetero-fructooligosaccharides and hetero-levans [16]. Multiple disaccharides can also be utilized as acceptor molecules, such as lactose, producing the prebiotic lactulose [18]. Non-carbohydrate acceptor molecules in LS-catalyzed transglycosylation reactions can be used to generate a variety of compounds. Mena-Arizmendi et al. [19] fructosylated aromatic compounds as well as aliphatic alcohols using the LS from *B. subtilis* [19]. Lu et al. [20] successfully, albeit in low yields, fructosylated isopropanol and 1-pentanol, while larger alcohols were unsuccessful as acceptors with LS from *Bacillus licheniformis*. The authors suggested that this lack of success was due to the increased hydrophobicity of the larger alcohol acceptors [20].

Enzymatic β-(2→6) transfructosylation is restricted by the low number of LS enzymes available [21]. Therefore, it is of interest to discover new LSs with a wide acceptor substrate specificity. Previously, genome mining was utilized to search for new LSs with novel properties [22]. After a screening and an examination of their thermal stability and kinetic parameters, the top candidates were identified, which included the LSs from *Beijerinckia indica subsp. indica*, *Paraburkholderia graminis*, *Vibrio natriegens*, *Novosphingobium aromaticivorans*, and *Gluconobacter oxydans*. The objective of this study was to investigate the end-product profiles of the reactions catalyzed by the selected LSs and to study their acceptor specificity using mono- and di-saccharides as well as two alcohols, an aliphatic alcohol and an aromatic one. The results of their product profiles were related back to the amino acid sequence of each enzyme, to connect the enzymatic structure with LS activity.

## 2. Results and Discussion

### 2.1. Time Courses for LS-Catalyzed Transfructosylation Reaction with Sucrose

The conversion rate of sucrose was analyzed (see Figures 1–5). For the LS *V. natriegens*, the consumption of sucrose was constant from the 4–12 h reaction, leading to a total conversion of 60% of the starting material by hour 50 (see Figure 1). The conversion rate of the sucrose decreased over the time course, which can be the consequence of a lower sucrose concentration and/or product inhibition. The rate of sucrose consumption by LS from *G. oxydans*, *N. aromaticivorans*, *P. graminis*, and *B. indica subsp. indica* was constant throughout the initial stage of the reaction, with final consumptions of 78%, 53%, 81%, and 72%, respectively, at 50 h. The LSs from *G. oxydans* and *P. graminis* almost depleted the sucrose as seen in Figures 2 and 3. Subtracting the concentration of free fructose from that of glucose provides the transfructosylation extent of the LS-catalyzed reaction to produce levan and FOSs. Although transfructosylation products are formed, the concentration of glucose is more or less close to that of fructose in the product profiles of the LS from *V. natriegens* (56 g/L glucose, 45 g/L fructose at 50 h) and *N. aromaticivorans* (35 g/L glucose, 27 g/L fructose at 50 h) throughout the course of the reaction. There is the possibility that glucose is consumed and directed towards the formation of blastose [23].

The extent of the transfructosylation reaction in the presence of LS from *V. natriegens* increased to 40% at 8 h. When the reaction was stopped at 50 h, transfructosylation decreased to 20%; this decrease was in accordance with the hydrolysis of some end-products. The ratio of transfructosylation products over hydrolytic products in the reaction catalyzed by LS from *V. natriegens* decreased from 5.1 to 0.6 over the reaction time course. This decrease was mainly due to the release of monosaccharides; indeed, the amount of transfructosylation products only decreased from 78 g/L to 63 g/L, while the monosaccharides content increased from 15 g/L to 100 g/L. The high production of hydrolysis products at 50 h corresponded with the kinetic parameters of the LS from *V. natriegens*, where both the turnover rate ($k_{cat}$) and the catalytic efficiency were higher for hydrolysis (246 $s^{-1}$, 101 $s^{-1} \cdot mM^{-1}$) than they were for transfructosylation (152 $s^{-1}$, 0.350 $s^{-1} \cdot mM^{-1}$) [22].

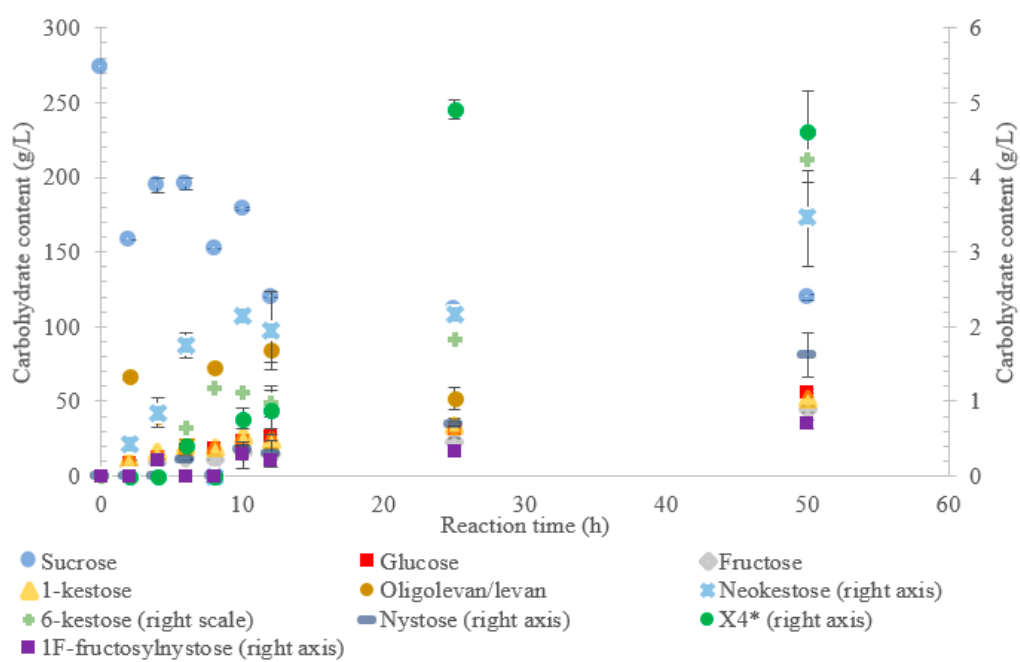

**Figure 1.** Reaction product profile of LS from *Vibrio natriegens* with sucrose for 50 h at 30 °C. *X4 is a tetrasaccharide peak separate from nystose.

Initially, within the first 2 h of incubation with sucrose, the reaction selectivity of LS from *G. oxydans* was dominated by transfructosylation. Transfructosylation decreased after this time, gradually increasing once again to parity with hydrolysis at 50 h. Comparing the transfructosylating products versus the monosaccharides produced, the ratio was in favor of transfructosylation during the first 6 h. It decreased to a ratio of 0.7 by 50 h. The amount of transfructosylating products increased from 71 to 88 g/L from 2 to 50 h, while the monosaccharide content increased from 21 to 125 g/L.

The results also show that hydrolysis was the dominant reaction for the LS from *P. graminis*, with transfructosylation products making up 39–46% of the products. In the initial stages of the reaction, similar amounts of monosaccharide products (25 g/L) and transfructosylation products (21 g/L) were generated; as time progressed, the monosaccharide content increased to 144 g/L and the transfructosylation products increased to 93 g/L. These results fit into the context of the turnover rates, with the transfructosylation turnover (4.72 $s^{-1}$) rate similar to that for hydrolysis (5.76 $s^{-1}$), resulting in an initial high production of transfructosylation products, dropping due to the higher catalytic efficiency of hydrolysis (0.289 $s^{-1} \cdot mM^{-1}$) as compared to transfructosylation (0.00987 $s^{-1} \cdot mM^{-1}$) [22].

Similarly, the reaction catalyzed by the LS from *N. aromaticivorans* was dictated more by its hydrolytic activity, with the amount of fructose used for transfructosylation reaching a maximum of 33%. While the hydrolytic turnover (176 s$^{-1}$) was similar to that of the transfructosylating rate (157 s$^{-1}$), the catalytic efficiency was much higher for hydrolysis (40.7 s$^{-1}$·mM$^{-1}$) than transfructosylation (0.303 s$^{-1}$·mM$^{-1}$) [22]. While the amount of free fructose caused by hydrolysis was greater than that used for transfructosylation, the amount of transfructosylated products was greater than the quantity of monosaccharides. The ratio between the transfructosylation over hydrolysis products started at 3.1 then decreased to 1.3 over the course of the 50 h. The amount of transfructosylated products increased from 32 to 83 g/L while the monosaccharides content increased from 10 to 62 g/L.

The transfructosylating activity of LS from *B. indica subsp. indica* increased over the 50-h reaction time, with the amount of fructose used for transfructosylation increasing to 45% by 50 h. Similar to LSs from *G. oxydans* and *P. graminis*, the LS from *B. indica subsp. indica* initially produced more transfructosylation products at 2 h (48 g/L) than monosaccharides (16 g/L). By hour 50, the monosaccharide content increased to 121 g/L, while the amount of transfructosylation products increased to 80 g/L. Indeed, both the turnover rate for transfructosylation (9.19 s$^{-1}$) and hydrolysis (7.98 s$^{-1}$) were relatively similar for LS from *B. indica subsp. indica*, but the catalytic efficiency was much higher for hydrolysis (0.382 s$^{-1}$·mM$^{-1}$) than for transfructosylation (0.020 s$^{-1}$·mM$^{-1}$) activities [22].

Figure 1 indicates that LS from *V. natriegens* produced levan-type oligo/polysaccharides (85 g/L; 76% *w/w* transfructosylation products) at 12 h (maximum oligolevan/levan production), and 18% oligosaccharide (24%, *w/w* total transfructosylation products) made of a mixture of trisaccharides ($X_3$), tetrasaccharides ($X_4$), and pentasaccharides ($X_5$). Larger oligosaccharides ($X_8$ and $X_{11}$) were included in the oligolevan/levan quantification. The production of $X_5$, $X_8$, and $X_{11}$ was detected after 2 h, while the production of tetrasaccharides was only detected after 8 h. The later detection of tetrasaccharides can be due to a lack of their accumulation, resulting from their use as precursors for larger molecules, i.e., oligolevan/levan. In Figure 1, the quantity of oligolevan/levan is seen as increasing then decreasing. The larger oligosaccharides ($\geq X_6$) included within the quantity of oligolevan/levan were likely used as fructosyl donor molecules, varying the quantity of oligolevan/levan at any given time. 1-kestose was the predominant trisaccharide produced, with there being 12-times less 6-kestose. The lack of accumulation of 6-kestose can be due to its recapture by the enzyme and being utilized as a substrate for the generation of other oligosaccharides and oligolevan/levan. This is what occurred when sucrose was incubated with the LS from *G. diazotrophicus* SRT4 [23]. LSs have demonstrated the ability to produce similar amounts of 1-kestose as inulosucrases from the same bacteria [12].

The incubation of the LS from *G. oxydans* with sucrose over 50 h (Figure 2) produced predominately oligolevan/levan (37%, *w/w* transfructosylation products), neokestose (49% *w/w*), and tetrasaccharides (9.0% *w/w*). Among the trisaccharides, 1-kestose accumulation matched that of 6-kestose, while neokestose was produced in large amounts. The inulin-type glycosidic linkage found on 1-kestose and neokestose may have made it difficult for further transfructosylation, which is supported by the low amounts of nystose and the accumulation of neokestose. High oligolevan/levan production (82% *w/w* at 2 h) corresponds with the results previously reported in [22]. This high-level oligolevan/levan production reveals that the LS from *G. oxydans* predominately performs a processive reaction, retaining the product in the active site, ready to accept more fructosyl residues.

The LS from *P. graminis* used the highest amount of sucrose amongst all the enzymes tested, as shown in Figure 3. The high bioconversion of sucrose by LS from *P. graminis* may be attributed to its high stability and/or to the low substrate/product inhibitions. Indeed, this LS was found to have very high thermal stability, with a half-life of 291 min at 50 °C [22]. There were less peaks indicating larger oligosaccharides as compared with the other LSs in this study, with the largest identified as $X_7$, while the highest quantity product was 1-kestose (74% *w/w* transfructosylation products), with no levan produced. In the first 2 h, no evidence of 6-kestose was observed, while there was an accumulation of 1-kestose, levan, and $X_4$. Again, 6-kestose was likely utilized as an acceptor substrate, converted into larger products, such as $X_4$ and levan. This enzyme had more flexibility in the type of transfructosylation reaction catalyzed with almost half of the tetrasaccharides produced from the inulin-type nystose (6.0 g/L), and fructosylnystose (5.1 g/L, 5.5% *w/w* transfructosylation products) was synthesized to a good degree.

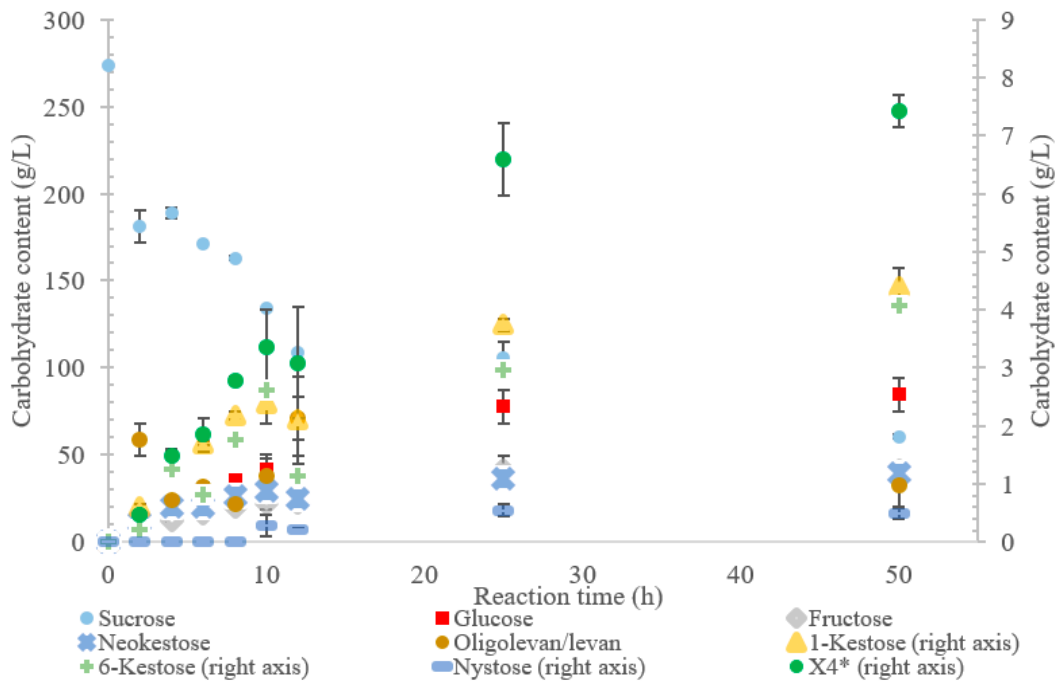

**Figure 2.** Reaction product profile of LS from *Gluconobacter oxydans* with sucrose for 50 h at 30 °C. *$X_4$ is a tetrasaccharide peak separate from nystose.

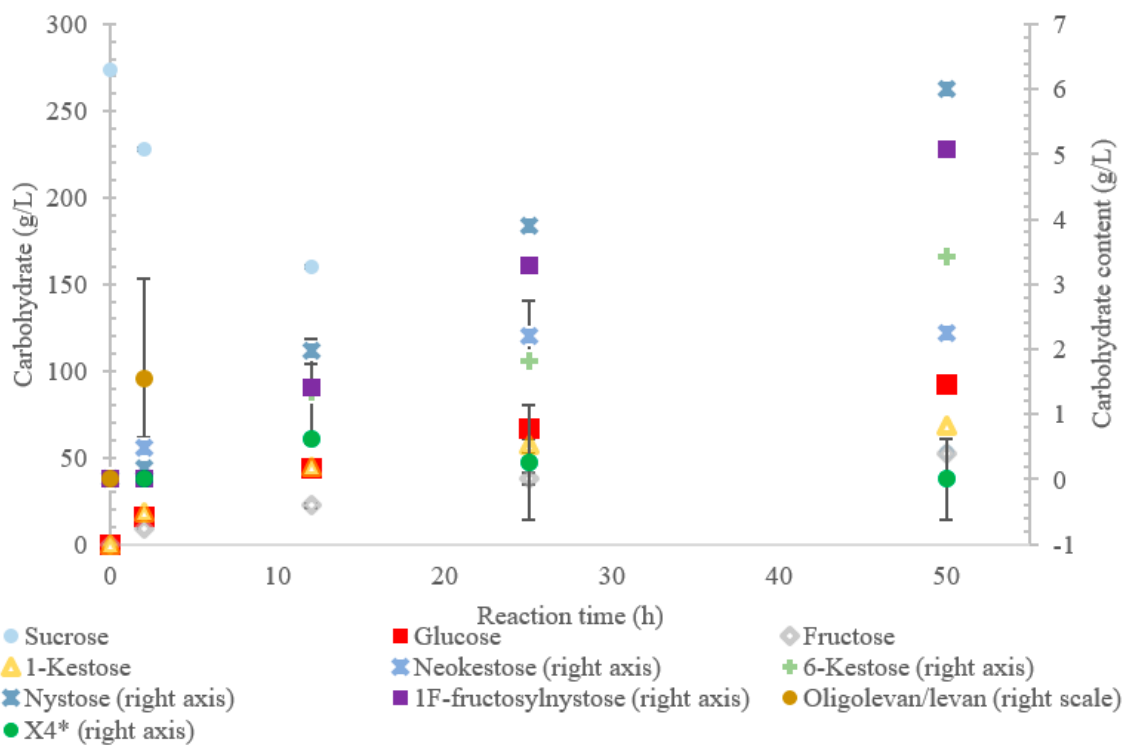

**Figure 3.** Reaction product profile of LS from *Paraburkholderia graminis* with sucrose for 50 h at 30 °C. *$X_4$ is a tetrasaccharide peak separate from nystose.

The LS from *N. aromaticivorans* utilized the least amount of sucrose after 50 h, with oligolevan/levan production increasing from 25 to 50 h (45%, *w/w* transfructosylation products) and total oligosaccharides increasing from 2 to 25 h, then plateauing between 25 and 50 h (51%, *w/w* transfructosylation products) (Figure 4). The main products were 1-kestose, 6-ketose, and $X_3$, $X_8$, $X_4$, $X_{13}$, and $X_{11}$. Production of 1-kestose and 6-kestose was linear until the 10-h reaction time (19 and 5 g/L), while the production of neokestose was linear until 25 h. By hour 50, more $X_3$ was synthesized than the other oligosaccharides. There was an accumulation of 1-kestose (36%, *w/w*) and 6-kestose (9.1% *w/w*), with only a small portion of $X_3$ products being comprised of neokestose (1.4% *w/w*). The lower amounts of neokestose in comparison to 1-kestose and 6-kestose, coupled with the production of larger oligosaccharides, provide evidence of a more processive reaction, where the product is retained within the active site rather than released after transfructosylation.

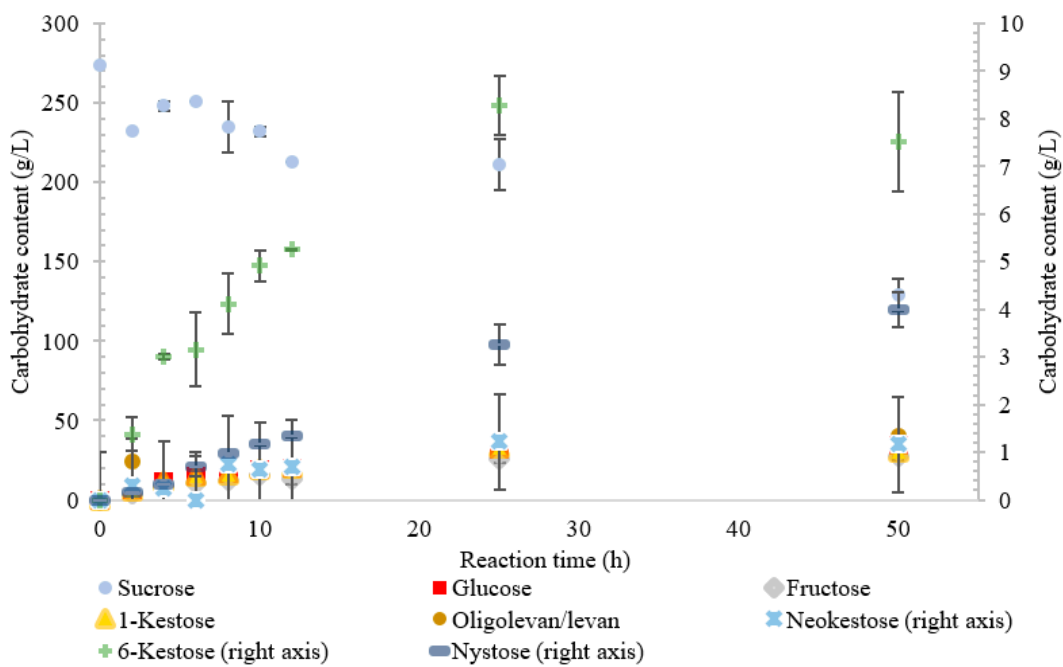

**Figure 4.** Reaction product profile of LS from *Novosphingobium aromaticivorans* with sucrose for 50 h at 30 °C.

Oligolevan/levan production from LS from *B. indica subsp. indica* started high (38 g/L, 79% *w/w* transfructosylation products), with the quantity decreasing between 4 and 50 h (24 g/L, 34% *w/w*) as seen in Figure 5. During this time, the amount of transfructosylating products steadily increased (48 to 77 g/L), with the main products (Figure 5) consisting of glucose, fructose, $X_3$, and $X_4$, with a few peaks representing larger oligosaccharides $X_5$ and $X_7$. During the 50-h incubation time, there was an increased production of 1-kestose (17–40% *w/w*) and neokestose (4–7% *w/w*), with a small accumulation of 6-kestose (0.9–3%).

The oligolevan/levan profiles of the LSs from *V. natriegens* (67 g/L at 2 h; 84 g/L at 12 h), *G. oxydans* (82 g/L at 2 h, 71 g/L at 12 h), and *B. indica subsp. indica* demonstrated high production followed by a decrease. These LSs may have demonstrated exolevanase activity once the oligolevan/levan accumulation was high enough to be hydrolyzed in the presence of the preferred substrate sucrose. This exolevanase activity was previously demonstrated by BS SacB (LS from *B. subtilis*), where hydrolysis (without sucrose present) continued until β(2-1) was reached [24]. In the presence of sucrose, these hydrolysis products then have the potential to be used as acceptor molecules in transfructosylation reactions.

Comparing the spectra of the five enzymes, the LS from *G. oxydans* and *N. aromaticivorans* produced the largest oligosaccharides (oligolevans) of up to 13 residues produced by *G. oxydans* and *N. aromaticivorans* LSs', followed by the LS from *V. natriegens* ($X_{11}$). The LS from *P. graminis* (74 g/L) produced 51–20% more trisaccharides as compared to the other enzymes studied. Comparing the production of FOSs from $X_3$–$X_5$ by weight, the LS from *P. graminis* produced the most at 93 g/L, followed by the *V. natriegens* (66 g/L), *G. oxydans* (56 g/L), *B. indica* subsp. *indica* (53 g/L), and lastly *N. aromaticivorans* (43 g/L). The order changed slightly when comparing the total transfructosylation products (including levan). The largest amount of total transfructosylation products was produced by the LS from *P. graminis* (93 g/L), *G. oxydans* (88 g/L), *N. aromaticivorans* (83 g/L), *B. indica* subsp. *indica* (77 g/L), followed lastly by *V. natriegens* (58 g/L). The highest FOS yields percentage wise, as compared to the initial amount of sucrose, all occurred at 50 h, with the exception of the LS from *N. aromaticivorans*, whose oligosaccharide yield plateaued at 25 h. The highest was achieved by LS from *P. graminis* (34%), which is high compared to *B. macerans* EG-6 (33%) and *Z. mobilis* (32%) [25,26]. LS from *V. natriegens* produced the most levan (85 g/L), followed closely by *B. indica* subsp. *indica* (84 g/L), then by *G. oxydans* (71 g/L), *N. aromaticivorans* (40 g/L), and *P. graminis* (2.0 g/L). The yields from both *V. natriegens*, *B. indica* subsp. *indica*, and *G. oxydans* are all quite high for Gram-negative bacteria. Their levan production resembles LS from Gram-positive bacteria, such as *B. subtilis* CCT7712 (111.6 g/L) and *B. methylotrophicus* SK 21.002 (100 g/L) [27,28]. LS of *N. aromaticivorans* resembles a more typical Gram-negative levan production, similar to that *G. diazotrophicus* SRT4 [29], with levan production of 24.7 g/L or *A. xylinum* NCIM 2526 with the production of 13.25 g/L [30].

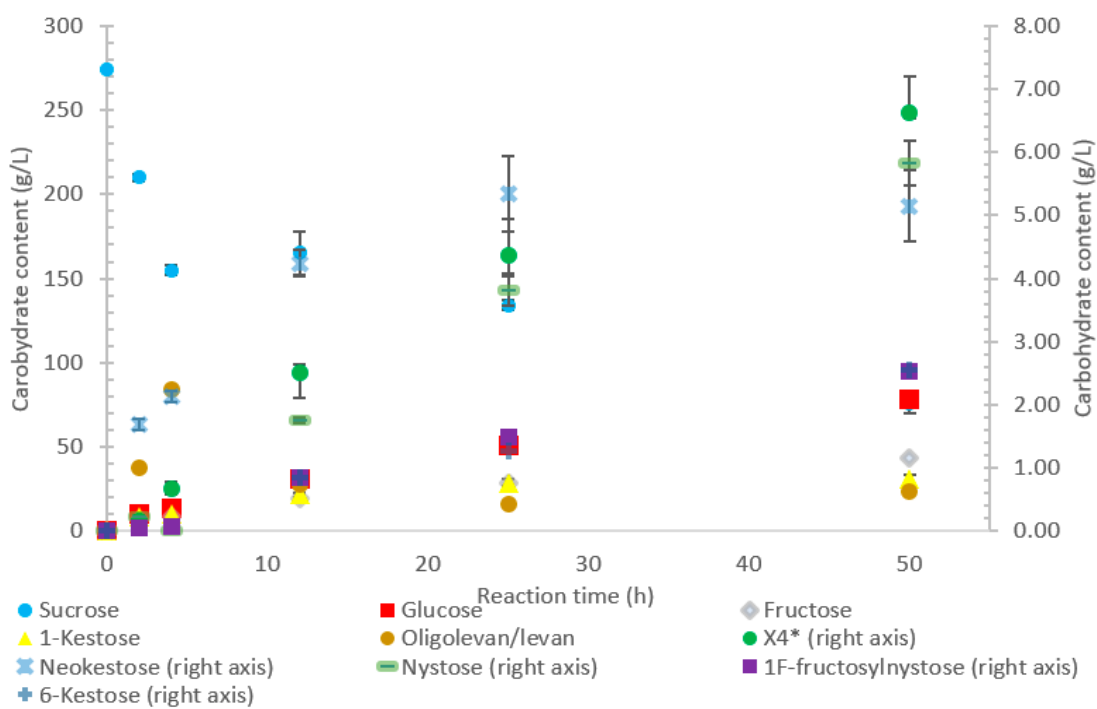

**Figure 5.** Reaction product profile of LS from *Beijerinckia indica* subsp. indica with sucrose for 50 h at 30 °C. *$X_4$ is a tetrasaccharide peak separate from nystose.

*2.2. Amino Acid Sequence Comparison*

The amino acid sequence of each of five LS enzymes were compared against the sequences belonging to LS to which there are crystal structures available. These sequences include LS from *B. megaterium* (PDB ID 3OM2, mutation D257A), *B. subtilis* (PDB ID 1OYG), *E. amylovora* (PDB ID 4D47), and *G. diazotrophicus* (PDB ID 1W18). The amino acid sequence of LS from *B. amyloliquefaciens*, whose characteristics have been extensively studied in our lab, was also added to the alignment. The sequence-based alignment of all the LSs are shown in Figure 6. Clustal Omega was used to perform the alignment and the results were analyzed using Jalview 2.10.1 [31,32]. The three amino acids essential to catalysis, Glu342, Asp86, and Asp247, from the LS from *B. subtilis* (Bs SacB) were conserved amongst all LSs [33].

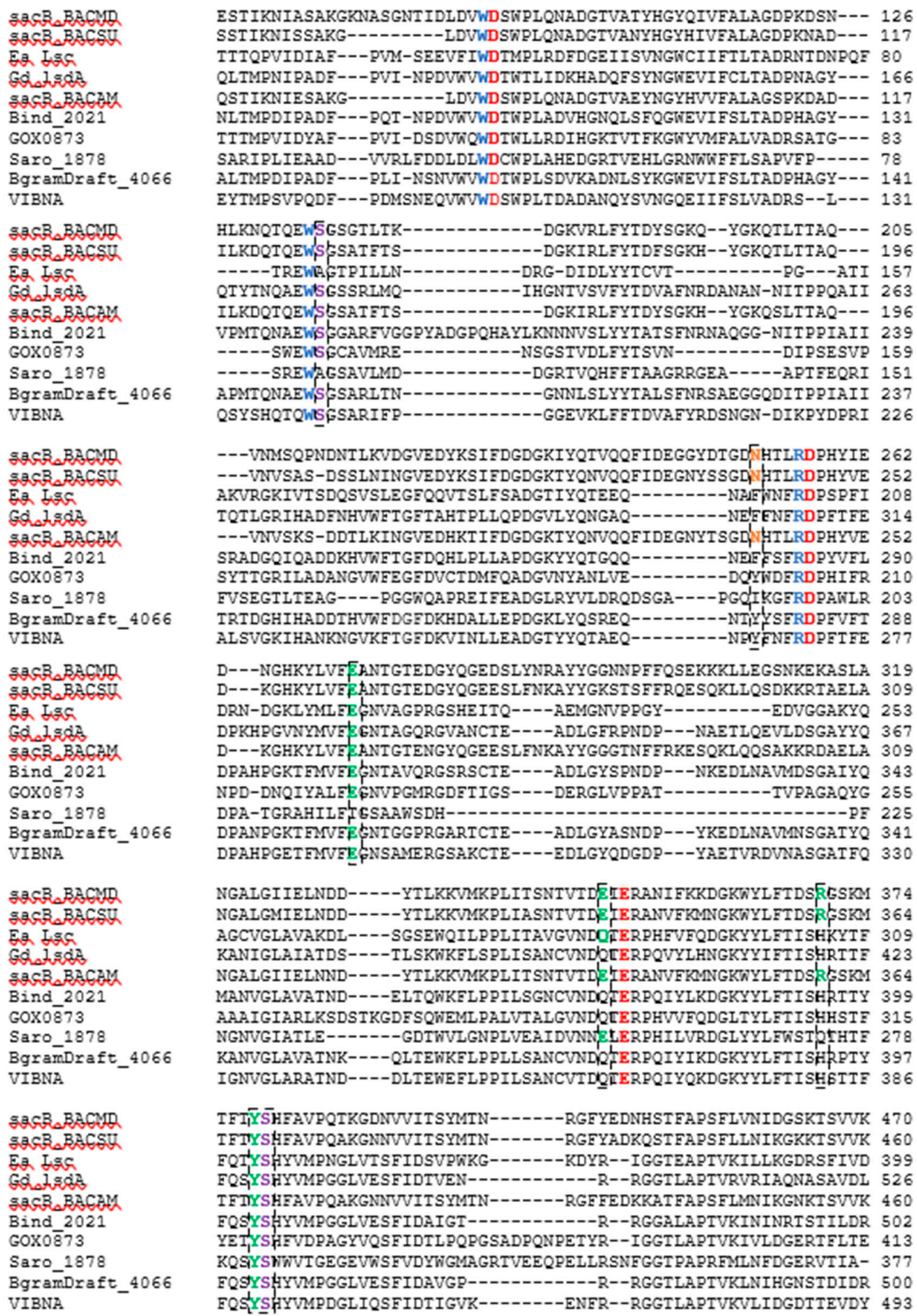

**Figure 6.** CLUSTAL O (1.2.4) multiple sequence alignment of LS enzymes with characterized crystal structures and enzymes investigated. The sequences for the LSs from *B. megaterium* (SacB_BM, D5DC07), *B. subtilis* (SacB_BS, P05655), *E. amylovora* (Ea Lsc, A0A0M3KKU6), *G. diaztrophicus* (Gd lsdA, Q43998), *B. amyloliquefaciens* (SacB_BACAM, E1UUH6), *B. indica* subsp. *indica* (Bind_2021, B2IF78), *G. oxydans* (GOX0873, Q5FSK0), *N. aromaticivorans* (Saro_1879, Q2G754), *P. graminis* (BgramDRAFT_4066, B1G3 × 6), and *V. natriegens* (VIBNA, A0A0S3EPZ1). Some residues excluded.

When a donor sucrose enters the active site, it is oriented and positioned by amino acids corresponding to the −1 and +1 subsites. Trp85, Arg246, and Trp163 from Bs SacB; and Trp95, Arg256, and Trp172 from Bm SacB (LS from *B. megaterium*) are highly conserved within family GH68 and contribute to the −1 subsite, which interacts with the fructosyl unit of sucrose [34]. These residues were entirely conserved in the five examined LSs. Other amino acids (Bs SacB) that interact with the incoming sucrose at the glucosyl residue are Glu340, which forms H-bonds with 3-OH and 4-OH; Arg360 interacts with 2-OH and 3-OH groups; Tyr411 interacts with 2-OH; and Arg246 forms H-bonds with 4-OH of glucopyranoside residue as well as 3-OH of the fructofuranoside group [35,36]. The investigated residues corresponding to Tyr411 and Arg246 were both conserved, while Glu340 is maintained for LS from *B. amyloliquefaciens* and *N. aromaticivorans* but was replaced with glutamine in LSs expressed by *B. indica subsp. indica*, *G. oxydans*, *P. graminis*, and *V. natriegens*. LS from *E. amylovora* and *G. diazotrophicus*, both Gram-negative LSs, had a glutamine residue at this position [11,14]. The sucrose-binding site was found to be maintained with this substitution [11]. In the Bs SacB, two serine residues (Ser164 and Ser412) formed H-bonds with Asp86, orientating it within the active site. Ea Lsc (*E. amylovora*) only had one serine instead of two (Ser353), with the other serine replaced by an alanine [14]. Hommann et al. [35] found that a mutation of this serine, Ser173 in Bm SacB to an alanine, did not alter the binding mode of sucrose but did decrease catalytic activity. The LS from *N. aromaticivorans* contained an Ala118 and a Ser323. All the other enzymes contained serine residues at both positions.

Glu262 from BS SacB is part of the hydrogen bond network along with Arg246, Tyr411, Arg360, and Glu342 [33], and was conserved amongst all LSs currently examined, with the exception of the LS from *N. aromaticivorans*, which had a Thr215 at this position. This hydroxyl group will potentially form hydrogen bonds, but the length of the side chain is slightly smaller than that from glutamate. This may decrease the intensity of the hydrogen bond network, perhaps increasing the flexibility of the active site.

Arg360 in BS SacB, essential for polysaccharide synthesis, was replaced with His in *B. indica subsp. indica*, *G. oxydans*, *P. graminis*, and *V. natriegens* [33]. This substitution is typical with Gram-negative bacteria and leads to the formation of oligosaccharides instead of levan [14]. Surprisingly, *G. oxydans* produced a large amount of levan, while *V. natriegens* and *P. graminis* produced smaller amounts as seen in Figures 1–3. This residue in *N. aromaticivorans* LS was replaced by a Gln274. While glutamine has a similarly polar side chain, it is longer than that of arginine.

Another amino acid, which was found to be essential for polysaccharide growth, was Asn242, located in subsite +2 [33]. In this position (Asn252), a mutation to alanine in *B. megaterium* SacB removes polymerase activity, while a mutation to an amino acid with a side chain, such as aspartate, retained polymerase activity [35]. This residue was conserved for LS from *B. amyloliquefaciens*. At this position, *N. aromaticivorans* (Ile193), *B. indica subsp. indica* (Phe280), *P. graminis* (Tyr278), and *G. oxydans* (Tyr200) replaced the asparagine with other residues. An analysis of enzyme–carbohydrate interactions of 60 different glycosyltransferases and glycosylhydrolases by Charoenwongpaiboon et al. [37] revealed that 80% of the residues interacting with the sugars were charged and polar. Isoleucine is neither charged nor polar, indicating no interactions with this residue and the incoming carbohydrates. This asparagine was maintained in both *E. amylovora* (Asn200) and *G. diazotrophicus* (Asn245) but was located 10 Å away and was unlikely to interact with the saccharide [14]. The asparagine in *V. natriegens* (Asn269) aligned with that of *E. amylovora* and *G. diazotrophicus*.

The difference in the quantity and size of the FOS production may have been due to differences in a pocket, which has been shown to retain acceptor molecules. Polsinelli et al. [38] determined that there was a levanbiose-binding site on LS from *Erwinia tasmaniensis* and *E. amylovora*, defined by Ala34, Phe35, Pro36, Val37, Arg73, Ile89, Trp371, Phe376, Arg377, and Ile378 (residues correspond to *E. amylovora*). The authors suggested that the binding of levanbiose or another small oligofructan to the enzymatic structure close to the active site would increase its likelihood to be used as an acceptor molecule, thereby increasing FOS production [38]. There were substitutions in each of the five enzyme sequences, with LS from *P. graminis* containing smaller and less polar residues (Ala79Asp, Val37Leu, Arg73Pro, Ile89—, Trp371Pro, and Phe376—). These changes may have reduced the affinity for small levan-type oligosaccharides, making sucrose the most used acceptor molecule, producing larger quantities of the trisaccharide 1-kestose (74% *w/w* transfructosylation products, 68 g/L). Arg377, which has the potential to form three hydrogen bonds with levanbiose (as well as other acceptors), was maintained in all the enzymes tested.

*2.3. Examination of LS Active Site*

After a dedicated search in the Protein Data Bank, different structural templates for each LS were selected, sequence alignments were controlled manually after precise checking of the quality of the structures, 500 (5 × 100 each) structural models were generated, and the best structural model was selected based on the lowest DOPE score (see the material and methods). A homology-based model of LS from *B. amyloliquefaciens* was created using the crystal structure of LS from *B. subtilis* (PDB ID 1PT2), bound with sucrose with the mutation E342A. Homology models were created for LS from *B. indica subsp. indica* and *P. graminis* from the crystal structure of *G. diazotrophicus* (PDB ID 1W18), *V. natriegens* from the crystal structure of β-fructofuranosidase by *Microbacterium saccharophilum* K-1 (PDB ID 3VSR), and the LS *G. oxydans* using the crystal structure of *E. amylovora* (PDB ID 4D47) with the sucrose hydrolysis products trapped within the active site. No model was able to be developed for the LS from *N. aromaticivorans*.

In a second step, Autodock 4.2 was used to perform rigid docking with fixed LS models and flexible ligands, sucrose, glucose, and fructose. The representation of the molecule within the active site represents the orientation with the least binding energy conformation. The models with the docking for each enzyme are represented in Figures 7–9, with additional information found in Table S1 and Figures S1–S3. Docking was used to examine some of the catalytic differences between each enzyme. The electrostatic surface potential of the models was used to observe potential interactions between the enzyme and the substrates.

Initial observations of the cavities of the enzymes tested and *B. amyloliquefaciens* LS were compared to the deep negatively charged pocket of Bs SacB (see Figure 7A). The shape of the active site of the LS from *B. amyloliquefaciens*, based upon the electrostatic surface potential, is similar to that of Bs SacB, with more charged residues. Comparing the least binding energy conformation of sucrose (see Figure 7B), the sucrose is turned counterclockwise by approximately 15° and inverted. The cavity of the LS for *B. indica subsp. indica* (see Figure 7C) was wider, with a deeper region and fewer charged residues. Sucrose was positioned much in the same way as Bs SacB, with the glucopyranoside residue slightly tilted backwards. The cavity of the LS from *G. oxydans* (Figures 7D and 8D) was similarly shaped but less deep, with many charged residues concentrated together on the exterior of the active site. Sucrose within the active site was rotated 45° downwards. The *P. graminis* LS active site (see Figure 7E) is deeper than that of Bs SacB. It has more positive electrostatic potential, with sucrose in a very similar, but less deep, orientation within the cavity. The *V. natriegens* LS active site (Figure 7F) was wider and shallower and had more positive and less negative electrostatic potential. Like *P. graminis* LS, the sucrose was orientated in the same way as Bs SacB.

The structures of these enzymes were analyzed in consideration of their catalytic behavior. The LS from *B. amyloliquefaciens*, a Gram-positive bacterium, produced dominantly high-molecular-weight levan up to $10^4$ kDa [39]. This LS was reported to consume 50% of its initial sucrose by hour 12, and 95% by 50 hours [18]. The rotation of sucrose found in the active site may have also contributed to this increased activity. While this LS has high transfructosylation activity, its hydrolysis activity was higher than it was for BS SacB. The extra charges within the active site may also contribute to this.

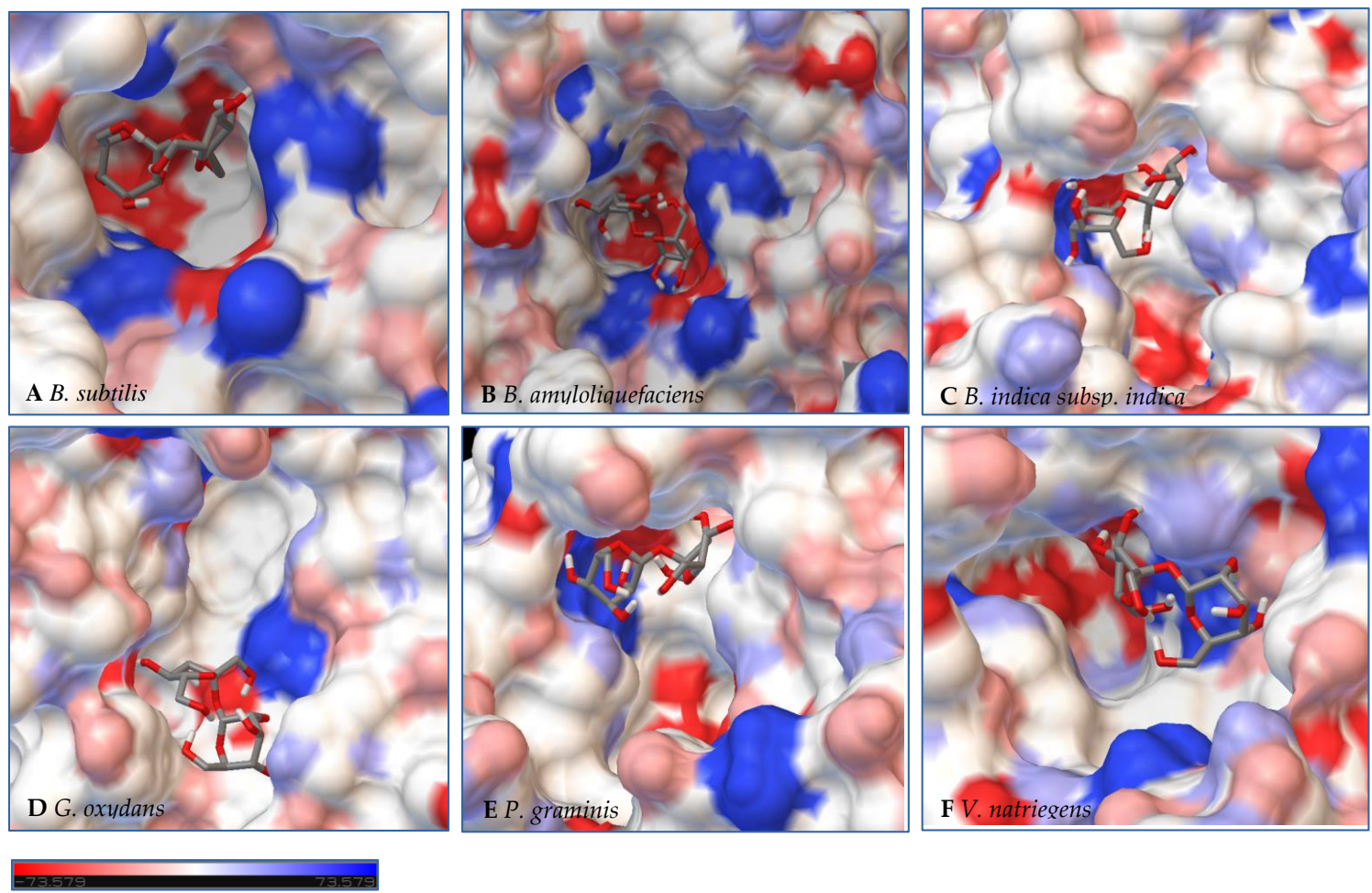

**Figure 7.** Models of LS from (**A**) *B. subtilis*, (**B**) *B. amyloliquefaciens*, (**C**) *B. indica subsp. indica*, (**D**) *G. oxydans*, (**E**) *P. graminis*, and (**F**) *V. natriegens* with sucrose docked within the active site. Sucrose is docked in a position representing the least binding energy state. Coloring represents the electrostatic surface potential ($K_bT/e_c$) with a spectrum of red (electronegative) to blue (electropositive). The surface charge was calculated using AutoDock v 4.2 and the PDBqt file was uploaded to OPAL webserver for PQR map and visualization was done using PyMol APBS plugin.

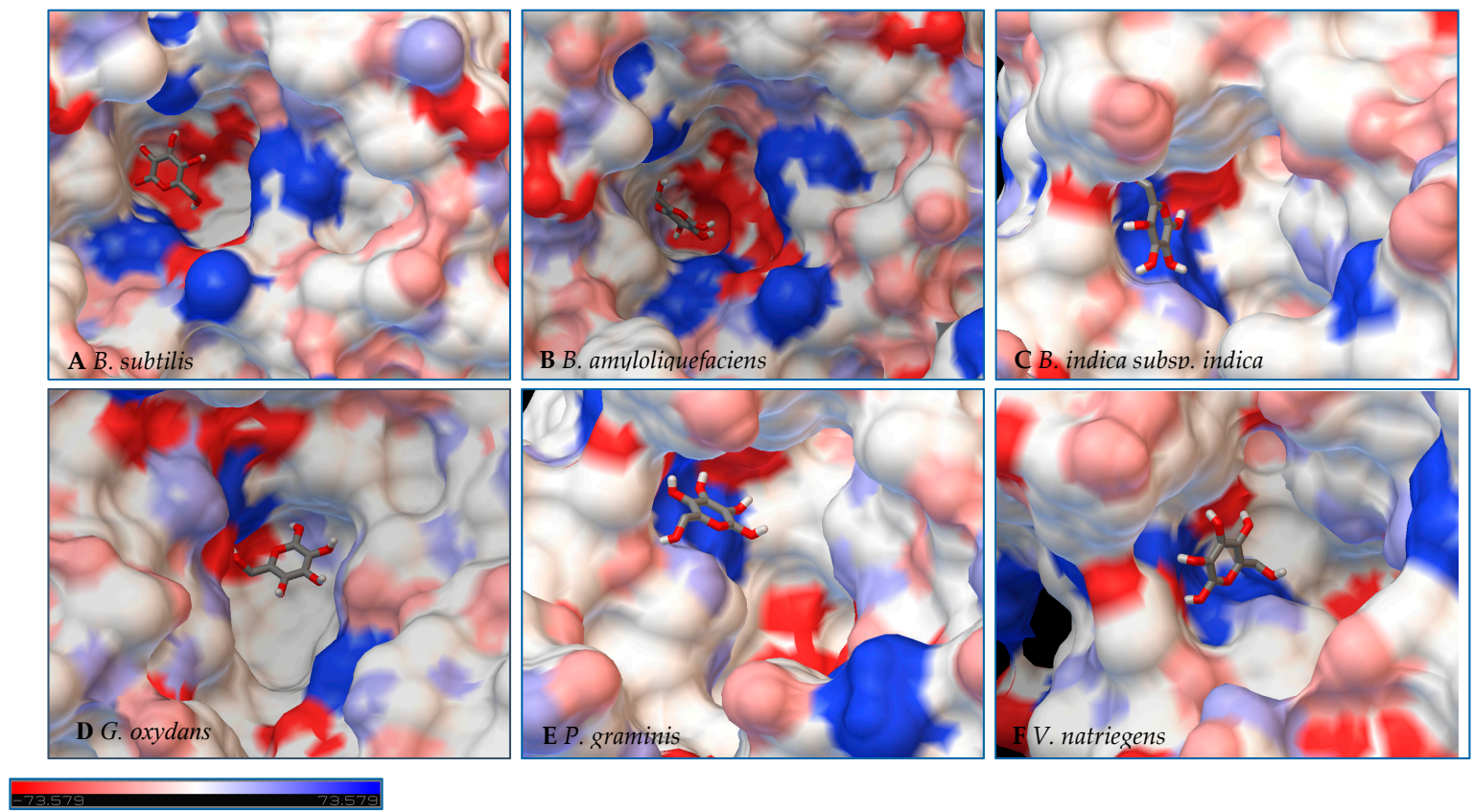

**Figure 8.** Models of LS from (**A**) *B. subtilis*, (**B**) *B. amyloliquefaciens*, (**C**) *B. indica subsp. indica*, (**D**) *G. oxydans*, (**E**) *P. graminis*, and (**F**) *V. natriegens* with glucose docked within the active site. Glucose is docked in a position representing the least binding energy state. Coloring represents the electrostatic surface potential ($K_bT/e_c$) with a spectrum of red (electronegative) to blue (electropositive). The surface charge was calculated using AutoDock v 4.2 and the PDBqt file was uploaded to OPAL webserver for PQR map and visualization was done using PyMol APBS plugin.

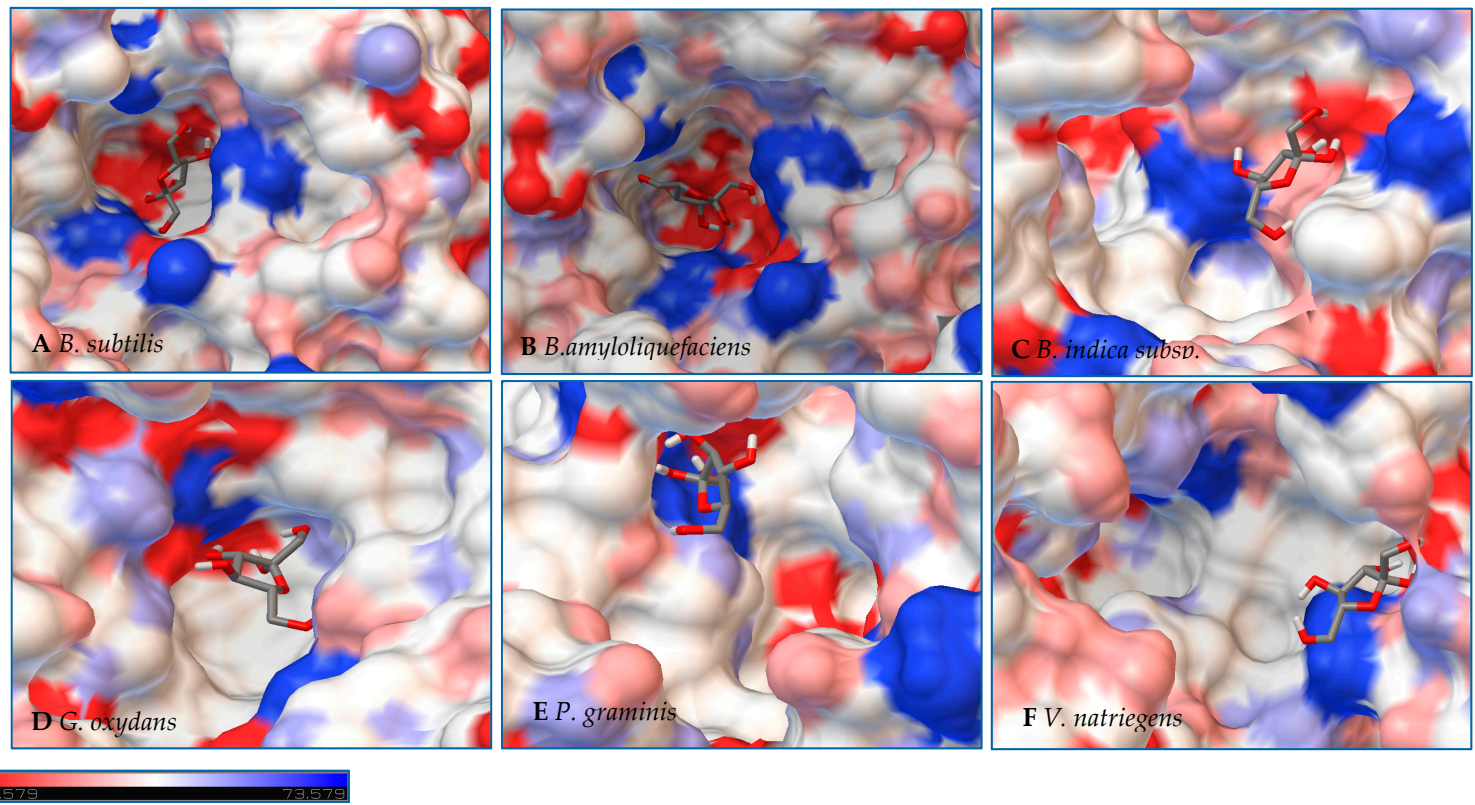

**Figure 9.** Models of LS from (**A**) *B. subtilis*, (**B**) *B. amyloliquefaciens*, (**C**) *B. indica subsp. indica*, (**D**) *G. oxydans*, (**E**) *P. graminis*, and (**F**) *V. natriegens* with fructose docked within the active site. Fructose is docked in a position representing the least binding energy state. Coloring represents the electrostatic surface potential ($K_bT/e_c$) with a spectrum of red (electronegative) to blue (electropositive). The surface charge was calculated using AutoDock v 4.2 and the PDBqt file was uploaded to OPAL webserver for PQR map and visualization was done using PyMol APBS plugin.

The interior of the LS from *G. oxydans* had a lack of charged residues as compared to Bs SacB, correlating to a lower affinity for sucrose, which along with the shallow pocket of the active site can result in faster turnover of the substrate, which was seen with the high sucrose consumption (78%). The concentrated residues on the exterior of the active site can interact with the growing fructan product, resulting in a processive reaction. The large levan produced by this enzyme (6986 kDa) and the high oligosaccharides (many peaks on the product spectra after tetrasaccharides) were evidence of this.

The LS from *V. natriegens* achieved a maximum ratio of 5.1 for transfructosylation versus hydrolysis. This enzyme also has a concentration of charged residues running along the left side of the active site (Figures 7F, 8F and 9F), helping to contribute to levan production, which was the dominant transfructosylation product. There were also smaller and fewer FOSs produced by the LS from *V. natriegens*. With the sucrose orientation within the active site, the products would be directed towards where the line of charged residues lie, contributing to the high initial level of transfructosylation.

The LS from *B. indica* subsp. *indica* also experienced high sucrose consumption, with catalysis initially dominated by transfructosylation, then by hydrolysis. Similar to *G. oxydans* LS, the cavity was wide but had fewer charged residues (Figures 7C, 8C and 9C). Levan dominated the transfructosylation products (79% *w/w*) initially, but by hour 50, most of the products resulted from hydrolysis. Unlike the LS from *G. oxydans*, there was no concentration of charged residues on the outer rim of the active site. The lack of charged residues close to the active site corresponds to the products of this LS, being of a low degree of oligomerization.

The LS from *P. graminis* experienced the highest consumption of sucrose (81%), but unlike the LS from *G. oxydans* and *B. indica* subsp. *indica*, the cavity was narrow in comparison. Initially, there was levan production, which was hydrolyzed by hour 50 and a resulting very high trisaccharide production. From Figures 7–9, all the residues reside in the same highly charged section of the pocket. The residues seen on the exterior of the pocket can contribute to stabilizing the polysaccharide as it grows, and possibly contribute to keeping it close to the active site for hydrolysis.

## 2.4. Acceptor Specificity of Selected LSs

LS' ability to use alternate acceptor molecules for transfructosylation can allow the synthesis of multiple sucrose analogs and FOSs headed with new groups. These molecules have a variety of potential applications, and a more simplistic route to their high-yield synthesis could be very advantageous. Due to the relaxed binding nature of the +1 subsite, there can be some variability of the docking of acceptor molecules [40]. Each LS examined was incubated with a fructosyl donor molecule (sucrose or raffinose) in excess and an acceptor molecule (sucrose, raffinose, glucose, galactose, maltose, lactose, xylose, sorbitol, and catechol). Comparing the percentage of each acceptor used for transfructosylation (Figure 10), the LS from *V. natriegens* consumed the combined largest variety of acceptor molecules, followed by LS from *P. graminis*, *G. oxydans*, *B. indica* subsp. *indica*, and *N. aromaticivorans*.

Comparing the use of sucrose and raffinose (Figure 10a–c, Table S2), LS from *V. natriegens*, *N. aromaticivorans*, and *B. indica* subsp. *indica* preferred raffinose as the substrate. LS from *Z. mobilis* was also found to prefer raffinose to sucrose as a fructosyl donor [41]. LS from *N. aromaticivorans* had the highest amount of hydrolysis (48%), compared with the other enzymes (16–38%). All LSs were able to produce a few different trisaccharides (1-3), tetrasaccharides (3-4), and pentasaccharides (1-3) using raffinose as the sole substrate. LS from *P. graminis* catalyzed the synthesis of multiple oligosaccharides using raffinose, a heptasaccharide ($X_7$), two octasaccharides ($X_8$), and a hendecasaccharide ($X_{11}$). There was less diversity of oligosaccharides, but each produced a hendecasaccharide and tridencasaccharides ($X_{13}$).

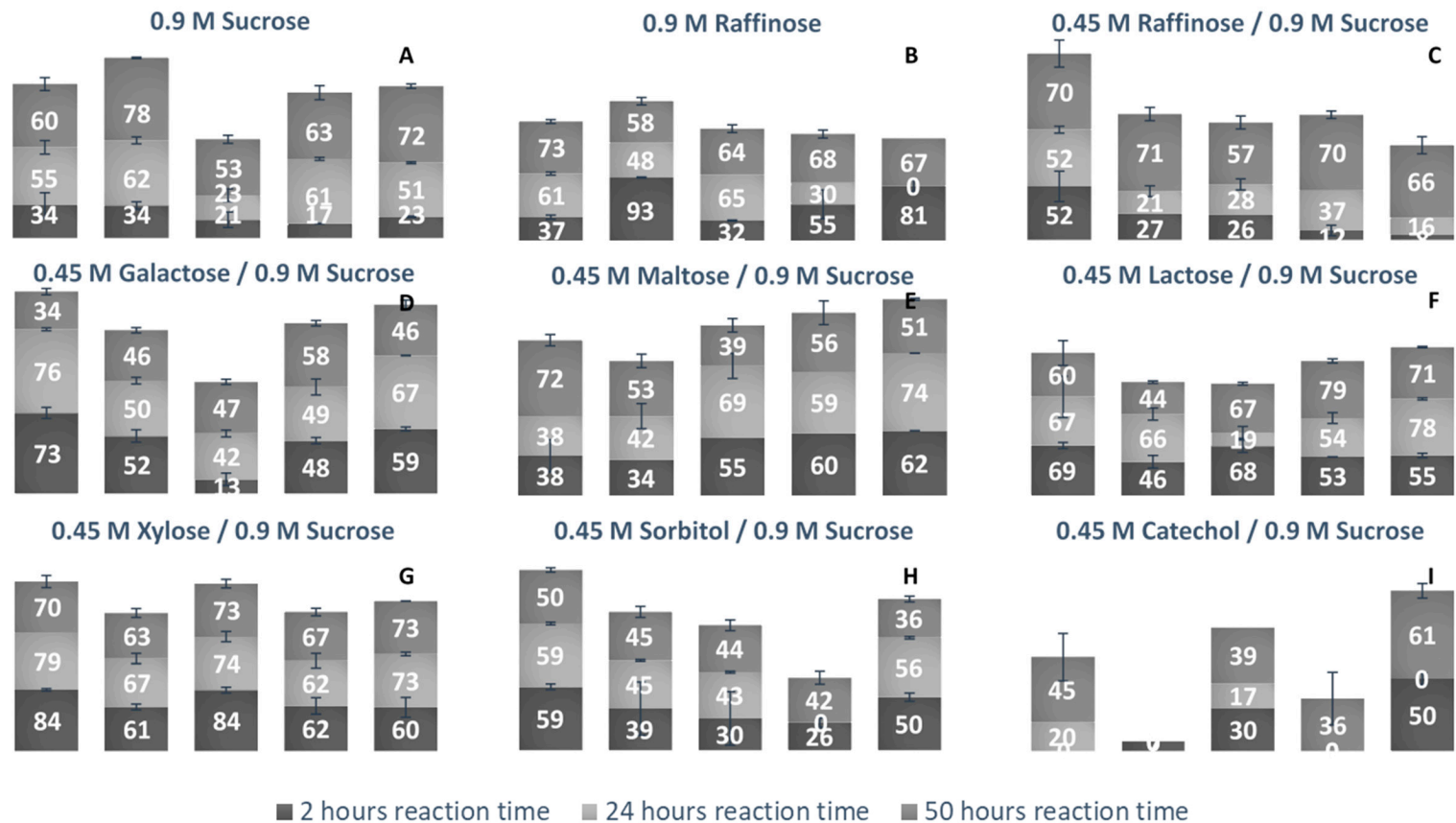

**Figure 10.** Percent bioconversion of various monosaccharides, disaccharides, and trisaccharides as acceptor/donor molecules by various LSs as a function of time. (**A**) Sucrose (0.9 M) was the sole fructosyl donor and acceptor initially present. (**B**) Raffinose (0.9 M) was the sole fructosyl donor and acceptor initially present. (**C**) Raffinose (0.45 M), in the presence of sucrose (0.9 M) were the initial substrates (both available to act as fructosyl donor/acceptor molecules). (**D**) Galactose, (**E**) Maltose, (**F**) Lactose, (**G**) Xylose, (**H**) Sorbitol and (**I**) Catechol (0.45 M) were present to act as a fructosyl acceptors in the presence of sucrose (0.9 M). Column 1: LS from *V. natriegens*; column 2: LS from *G. oxydans*; column 3: *N. aromaticivorans*; column 4: LS from *P. graminis*; column 5: *B. indica* subsp. *indica*. Numbers listed on columns represent the percentage of carbohydrate acceptor consumed at that given reaction time.

The synthesis of galactose-headed FOSs was desired since they can provide interesting prebiotic activity. Multiple species of LS can utilize galactose as a fructosyl acceptor molecule [21]. In Figure 10d, LS from *V. natriegens* consumed a large amount of galactose within the first 2 h of incubation, then by hour 50, much of the galactose was released into the system. LS from *B. indicia* subsp. indica and *G. oxydans* followed a similar trend. The release of galactose could be due to the hydrolysis of the fructosylated galactose products. When the sucrose analogue, Gal-Fru, was incubated with Bs SacB, 52% was converted into transfructosylated products, while the rest (48%) was hydrolyzed [16]. At 50 h, LS from *P. graminis* consumed the most galactose (58%). The greatest consumption of galactose occurred within the first 2 h of incubation, then slightly increased over the remaining 48 h. Early consumption of galactose by the LS from *V. natriegens*, *G. oxydans*, *P. graminis*, and *B. indica subsp. indica* infers that their Km values towards galactose values are relatively small.

All LSs tested here were able to consume maltose to a high degree. The utilization of maltose requires the transfructosylation of a glycopyranosyl residue, making the glycosidic bond between maltose and fructose similar to what occurs in sucrose. LS from *Rahnella aquatilis*, *B. subtilis*, *B. amyloliquefaciens*, and *Geobacillus steareothermophilis* were all able to use maltose to a high degree as an acceptor molecule [17,18,42,43]. The peaks produced using maltose and sucrose as the substrates were compared to those produced using solely sucrose. Two new trisaccharide peaks were identified from LSs from *B. indica subsp. indica,* while one new trisaccharide was produced by LSs from *N. aromaticivorans*, *G. oxydans*, *V. natriegens*, and *P. graminis*. Both *P. graminis* and *B. indica subsp. indica* catalyzed the synthesis of a pentasaccharide using maltose, while these enzymes along with *V. natriegens* were able to catalyze the synthesis of a heptasaccharide.

Another fructosyl acceptor, which was widely used by all LSs, was lactose. Each of these enzymes can be used to make the most desired prebiotic lactosucrose [7]. In all cases, lactose was predominately used within the first 2 h. With lactose, all enzymes were able to catalyze the synthesis of pentasaccharides except for the LS from *G. oxydans*, while this enzyme was able to produce a hexasaccharide using lactose. The LSs from *P. graminis* and *B. indica subsp. indica* catalyzed the synthesis of a heptasaccharide. All LS enzymes examined here (see Figure 10G) used xylose to a high degree within the first 2 h, with a minimum 60% consumed. LSs from *V. natriegens* and *N. aromaticivorans* had some of their xylose products hydrolyzed, returning xylose. Xylose was also found to be a good acceptor molecule for Bs SacB, yielding β-D-fructofuranosyl-α-D-xylopyranoside [17]. The reasoning for the high consumption of xylose is suspected to be due to the equatorial position of the C2-OH, which was found to be in a good position for protonation by the active site [16,33].

An alditol and a benzene diol were tested for their use as acceptor molecules. The ability to use alcohols as acceptor molecules is not a common attribute to LSs. Catechol had a much lower signal on the PAD, which made its detection difficult when quantities were low. The lowest detection limit of catechol was 10 µM, limiting its detection when the substrate was highly used. The LS from *V. natriegens* used the highest amount of sorbitol, most of it within the first 2 h. By hour 50, approximately 10% was hydrolyzed. The other enzymes used approximately 40% of the sorbitol. There was a peak indicating Sor-Fru and two new trisaccharide peaks from *N. aromaticivorans* LS and the *V. natriegens* LS. Additionally, a hexasaccharide peak was seen from *G. oxydans* and *B. indica subsp. indica*. Catechol was an efficient acceptor molecule for fructose. By hour 50, the LS from *B. indica subsp. indica* had consumed over 60% of the initial catechol. This enzyme produced new peaks, indicating a trisaccharide, a heptasaccharide, and a nonasaccharide. The LS from *G. oxydans* had a product profile indicating the use of catechol in a hexasaccharide and heptasaccharide while the LS from *P. graminis* catalyzed the formation of a heptasaccharide and a nonasaccharide using catechol. Mena-Arizmendi et al. [19] found that Bs SacB was more capable of fructosylating secondary alcohols than primary alcohols, due to an inverted relationship between pKa and the ability of Glu342 to deprotonate the hydroxyl group [19]. While the pKa of sorbitol is 13.57, and higher than the pKa of catechol (9.48), they were both similarly used as acceptor molecules [44]. The additional hydroxyl groups on sorbitol may have provided additional sites for stabilization within the active site, resulting in good yields for transfructosylation.

The *B. indica subsp. indica* showed a general trend of catalyzing the synthesis of the larger heteroFOSs (≥6 units). Using the example of the sucrose incubation, this enzyme also produced a large amount of levan (84 g/L) by hour 4 of the incubation, decreasing to 24 g/L by hour 50. With the acceptors and sucrose incubation, the levan may be produced with the acceptor molecule at its head and later hydrolyzed into the heteroFOSs. With the LS from *G. oxydans*, most frequently, the largest molecule incorporating the alternative acceptor molecule was a hexasaccharide. With sucrose as the sole substrate, this enzyme produced a large amount of hexasaccharides (5 g/L) as well as the smaller molecules (trisaccharides and tetrasaccharides). As seen in Figure 8D, there is a line of charged residues capable of interacting with the products. By comparing these results, these residues likely have a strong interaction up to six residues in length.

## 3. Materials and Methods

### 3.1. Enzyme Production, Recovery, and Purification

*Escherichia coli* BL21(DE3) cells (Invitrogen) were each transformed with the pET22b(+) expression vector containing the genes for: *Vibio natriegens* (VIBNA, A0A0S3EPZ1), *Gluconobacter oxydans* (GOX0873, Q5FSK0), *Novosphingobium aromaticivorans* (Saro_1879, Q2G754), *Paraburkholderia graminis* (BgramDRAFT_4066, B1G3X6), and *Beijerinckia indica* subsp. indica (Bind_2021, B2IF78).

A preculture, containing Lysogeny broth (LB) and ampicillin (100 µg/mL), was inoculated by selecting colonies grown on LB agar containing carbenicillin (1 µg/mL) and incubated for 24 h at 37 °C at 250 rpm. The preculture (2%) was used to inoculate the culture medium composed of terrific broth and ampicillin (100 µg/mL). The culture was grown at 37 °C at 250 rpm until growth achieved an optical density of 1.2 at 600 nm. Gene expression was then induced through the addition of β-D-thiogalactopyranoside (IPTG, 1 mM), with growth continued for 18 h afterwards at room temperature at 250 rpm. The cells were then collected by centrifugation (8000 rpm) and stored at −70 °C. Enzyme recovery was initiated by defrosting the pellet on ice for 40 min and resuspending the cells in sonication buffer (50 mM pipes, 300 mM NaCl, 10% glycerol *v/v*, pH 7.2). Lysozyme (4 mg/g pellet) and DNase (Sigma Chemical Co., St. Louis, MO, USA) (2000 U/mL) were added and the suspension was gently mixed at 40 rpm on ice for 30 min. The cells were lysed by ultrasonication using a microtip (Misonix Ultrasonic Liquid Processor S-4000, Farmingdale, NY, USA) for 1 minute (10 s on, 60 s off, amplitude of 15) in a salt-ice bath. The cellular debris was removed through centrifugation (45 min, 14,000 rpm, 4 °C). The supernatant was dialyzed against potassium phosphate buffer (5 mM, 12 L) at pH 6, then frozen, and lyophilized.

The crude enzyme extracts were resolubilized potassium phosphate buffer (50 mM, 0.5 mL) at pH 6.0, filtered, and loaded onto a Histrap FF (GE Healthcare, Mississauga, ON, Canada) 1 mL column. The column was washed with sonication buffer (15 mL), wash buffer (50 mM Pipes, 300 mM NaCl, 10% glycerol *v/v*, pH 6.4, 15 mL), 5 mM imidazole prepared in wash buffer (15 mL), 10 mM imidazole prepared in wash buffer (15 mL), and eluted with a gradient of imidazole (100 mM-200 mM, 3 mL each fraction) prepared in wash buffer. Each fraction was tested for activity by the 3,5-dinitrosalicyclic acid (DNS) activity assay. Active fractions were pooled together, dialyzed against potassium phosphate buffer (5 mM, 12 L), and lyophilized. Purity was confirmed by sodium dodecyl sulfate polyacrylamide gel electrophoresis (SDS-PAGE).

### 3.2. Time Course for LS-Catalyzed Reactions

LS (5-7 U/mL) was incubated with sucrose (0.8 M) at 30 °C in triplicate, with samples withdrawn at 2, 4, 6, 8, 10, 12, 24, and 50 h. The samples were boiled for 5 min to stop the reaction. To quantify the reaction products and the remaining sucrose, a sample (10 µL) of the reaction mixture was analyzed using high-pressure anion exchange chromatography (HPAEC) with a pulsed amperometric detector (PAD, Dionex/Thermo Fisher, Burlington, ON, Canada) using a Dionex ICS 3000 system (Dionex/Thermo Fisher, Burlington, ON, Canada) eluted on a CarboPac PA200 column (3 × 250 mm, Thermo Fisher, Burlington, ON, Canada). The sample was eluted with a linear gradient of 0–100% of 200 mM sodium acetate prepared in 100 mM NaOH for 25 min at a flow rate of 0.5 mL/min at 32 °C. Calibration standards of glucose, fructose, sucrose, 1-kestose, nystose, and $1^F$-fructosylnystose were used. Oligolevans were defined as larger oligosaccharides ($\geq X_6$) and were quantified from the remaining substrate and the quantifiable end-product concentrations.

*3.3. Acceptor Specificity*

LS (5-7 U/mL) was incubated with either sucrose (0.8 M), raffinose (0.8 M), or sucrose (0.8 M) and an acceptor molecule (0.4 M), including raffinose, glucose, galactose, maltose, lactose, xylose, sorbitol, and catechol. Samples were withdrawn at selected reaction times (2 to 50 h) and boiled for 5 min to stop the reaction. The reaction was characterized by HPAEC-PAD with the same conditions previously used on the CarboPac Pa200 column. They were also analyzed using a CarboPac PA20 column (Thermo Fisher, Burlington, ON, Canada), with samples eluted isocratically with 20 mM NaOH at a flow rate of 0.5 mL/min at 32 °C.

*3.4. Homology Modeling*

3.4.1. Peptide Identification

The LS from *B. amyloliquefaciens* was purified by size-exclusion chromatography and was separated by SDS-PAGE gel. The LS bands were removed and sent for proteomic analysis by mass spectrometry at the Plateforme de Protéomique—Centre de Recherche du CHU de Québec (Laval, Québec, Canada). The bands were subjected to tryptic digestion using a MassPrep liquid handling robot (Waters) following the protocol by Brotherton et al. [45]. After an initial reduction (10 mM dithiothreitol) and alkylation (55 mM iodoacetamide), the sample was digested with using 126 nM porcine trypsin (sequencing grade, Promega) at 58 °C for 1 h. The digestion products were extracted using 1% formic acid and 2% acetonitrile, which was then followed by 1% formic acid and 50% acetonitrile. The protein extracts were pooled, dried by vacuum centrifugation, and then resuspended into 10 μL of 0.1% formic acid for analysis by electrospray ionization mass spectrometry. The peptide samples were separated by an online reverse-phase (RP) nanoscale capillary liquid chromatography (nanoLC) and analyzed by electrospray mass spectrometry (ESI MS/MS) [46]. The fragments were analyzed using Scaffold software (version 4.0) (Proeome Software).

3.4.2. Homology Model Selection

Two iterations of PSI-BLAST (NIH) were used to mine the Protein Data Bank (rcsb.org) to search for sequences with the lowest e-value, maximum coverage, and highest sequence identity to the amino acid sequences of the LSs from *B. amyloliquefaciens*, *B. indica* subsp. *indica*, *P. graminis*, *G. oxydans*, *V. natriegens*, and *N. aromaticivorans*. Structural templates were selected from the homologous enzymes with crystal structures, which had the highest similarity to the LSs sequence (see Supplementary Materials, Source and Modeling Summary). The LS from B. subtilis (PDB ID 1PT2) [33], bound with sucrose with a E342A mutation, was chosen as the template for *B. amyloliquefaciens*; the LS *G. diazotrophicus* (PDB ID 1W18) was the template for *B. indica* subsp. indica and *B. graminis*; β-fructofuranosidase by *Microbacterium saccharophilum* K-1 (PDB ID 3VSR) [11,47]; and the LS from *E. amylovora* (PDB ID 4D47) with the sucrose hydrolysis products within the active site was the template for the LS from *G. oxydans* and *N. aromaticivorans* [14]. Figure S4 demonstrates the choice of the unliganed *Microbaterium saccharophilum* K-1 beta-fructofuranosidase (PDB 3VSR) over the fructose-bound template (3VSS). Sequence alignments were performed with MAFFT program (version 7.245) using G-INS-I option and observed with Jalview version 2 [32,48].

### 3.4.3. Homology Modelling

Structural templates that contained a substrate/ligand were stripped off their ligand to ensure confident modelling of the Apo-forms. The modelling was performed using MODELLER v9.17 after feeding a pairwise sequence alignment of the template structure to the LS of concern in pir format. The alignment was manually checked to ensure that essential amino acids are conserved. No restraints were imposed for the modelling. For each LS sequence of interest, 100 models were generated. A set of suitable models was selected based on the lower values of DOPE score (Discrete Optimized Potential Energy- translates to the fit of these models). These models were validated using ProSa-web Protein Structure Analysis [49,50]. TM-Align was used to align the selected set of models with their template structure in an all vs. all pairwise manner [51]. This generated a matrix that helped to identify the closest model to the template structure as well as the expanse of the model diversity. Using the TM-Score (closer to 1) and RMSD (closer to 0), a single model was selected to work further. Figure S5 shows an overlay of the models generated.

### 3.4.4. Docking Protocol

Docking was performed using Autodock 4.2.6 software as a local install with MGLTools [52]. Rigid docking was performed by fixing the Apo form LS models and keeping the substrates: Sucrose, glucose, and fructose, flexible to identify the best fit pose. From the *B. subtilis* LS crystal structure (PDB ID: 1PT2) that has sucrose bound, the binding pocket can be estimated. Thus, using *B. subtilis* as a control case, a grid space of 0.375 Å was selected around this pocket to probe the binding of flexible ligands. Each ligand, based on their stereochemistry, was given different torsional degrees of freedom [Glc:6, Fru:7, Suc:13]. Based on the 5,000,000 evaluations of the binding pose, the pose with the least binding energy was selected for each ligand. Docking experiments were used to evaluate the catalytic differences between the different analyzed enzymes on the selected substrates.

## 4. Conclusions

The investigated enzymes had interesting properties in terms of the production of high-level FOS. The LS *P. graminis* produced the largest quantity of FOSs (93 g/L), predominately inulin-type trisaccharides while two of these enzymes, the LS from *G. oxydans* and *N. aromaticivorans*, were both able to produce large oligolevans ($X_{13}$). A further approach to large FOS production may include the use of co-solvents. Co-solvents can aide in precipitating the larger product, thereby further driving the thermodynamic equilibrium towards FOS production.

Each enzyme investigated had a wide substrate specificity. A further examination of the reasons why this is can occur through docking experiments with the LS models, which is the next step in our study. The ability for these LSs to transfructosylate aromatic and aliphatic alcohols is an interesting attribute, providing more potential synthetic applications. Using substrate engineering, and potentially using protecting groups, the site of transfructosylation can be controlled [1]. With the success already achieved in these trials, there is the potential for the transfructosylation of more elaborate substrates for the easier synthesis of fructosylated compounds.

**Supplementary Materials:** The following are available online at http://www.mdpi.com/1422-0067/21/15/5402/s1. Table S1: Interactions between enzyme residues and carbohydrate. Figure S1. Models of LS from a) *B. subtilis*, (b) *B. amyloliquefaciens*, (c) *B. indica subsp. indica*, (d) *G. oxydans*, (e) *P. graminis* and (f) *V. natriegens* with sucrose docked within the active site. Figure S2: Models of LS from a) *B. subtilis,* (b) *B. amyloliquefaciens*, (c) *B. indica subsp. indica*, (d) *G. oxydans*, (e) *P. graminis* and (f) *V. natriegens* with glucose docked within the active site. Figure S3: Models of LS from a) *B. subtilis*, (b) *B. amyloliquefaciens*, (c) *B. indica subsp. indica*, (d) *G. oxydans*, (e) *P. graminis* and (f) *V. natriegens* with fructose docked within the active site. Figure S4: Comparison of the fructose docked in a position representing the least binding energy state generated by docking simulations using the unliganed (3VSR). Figure S5: Overlay of the homology models. Table S2. Percent bioconversion of various monosaccharides, disaccharides, and trisaccharides as acceptor/donor molecules by various LSs. Modeling summary also available as a supplemental material.

**Author Contributions:** Conceptualization, A.H. and S.K.; Methodology, A.H. and S.K., T.J.N. and A.G.d.B. (homology modeling); Data Curation, A.H. and T.J.N.; Formal Analysis, A.H., S.K and T.J.N. (homology modeling); Funding acquisition, S.K.; Investigation, A.H. and T.J.N. (homology modeling); Software, T.J.N. and A.G.d.B.; Supervision, S.K.; Validation, S.K. and A.H.; Visualization, A.H. and T.J.N. (homology modeling); Writing—Original Draft Preparation, A.H.; Writing—Review & Editing, S.K. and A.G.d.B. (modeling part). All authors have read and agreed to the published version of the manuscript.

**Funding:** This research was funded by The Natural Sciences and Engineering Research Council (NSERC) of Canada (#RGPIN-2016-05042), and financial infrastructure support from Canada Foundation for Innovation (CFI# 36708), is grateful acknowledged.

**Conflicts of Interest:** The authors declare no conflict of interest. The funders had no role in the design of the study; in the collection, analyses, or interpretation of data; in the writing of the manuscript, or in the decision to publish the results.

**Abbreviations**

| | |
|---|---|
| LS | Levansucrase |
| FOS | Fructooligosaccharide |
| HPAEC-PAD | High-pressure anion exchange chromatography with a pulsed amperometric detector |
| PDB | Protein data bank |
| DNS | 3,5-Dinitrosalicyclic acid |
| LB | Lysogeny broth |
| IPTG | β-D-thiogalactopyranoside |
| SDS-PAGE | Sodium dodecyl sulfate polyacrylamide gel electrophoresis |